\documentclass[10pt,showpacs,preprintnumbers,amsmath,amssymb]{revtex4}

\usepackage{graphicx}% Include figure files
\usepackage{bm}% bold math
\usepackage{color}
\def\slashchar#1{\setbox0=\hbox{$#1$}
   \dimen0=\wd0 \setbox1=\hbox{/} \dimen1=\wd1
   \ifdim\dimen0>\dimen1 \rlap{\hbox to \dimen0{\hfil/\hfil}} #1
   \else  \rlap{\hbox to \dimen1{\hfil$#1$\hfil}} / \fi}
\begin{document}
% \vspace*{4cm}
\title{Lepton production cross sections in quasielastic $\nu/\bar{\nu}-$A scattering}
\author{M. Sajjad Athar, F. Akbar, M. Rafi Alam, S. Chauhan, S. K. Singh and F. Zaidi}
\address{Department of Physics, Aligarh Muslim University, Aligarh-202 002, India}

\begin{abstract}
We present the results of (anti)neutrino induced CCQE cross sections from some nuclear targets in the energy
 region of $E_\nu \le 1~GeV$. The aim of the study is to confront electron and muon production
 cross sections relevant for $\nu_\mu \leftrightarrow \nu_e$ or $\bar\nu_\mu \leftrightarrow 
\bar\nu_e$ oscillation experiments. The effects due to lepton mass and its kinematic implications,
 second class currents and uncertainties in the axial and pseudoscalar form factors are discussed for (anti)neutrino induced reaction cross sections on
 free nucleon as well as the nucleons bound in a nucleus where nuclear medium effects influence the cross section. The calculations have been performed using 
 local Fermi gas model with nucleon correlation effects. The details are given in Ref.[1].
\end{abstract}
\maketitle

\vspace{-5mm}
\section{Introduction}
The recent measurement of the neutrino mixing angle $\theta_{13}$ in $\nu_\mu \to \nu_e$ oscillation experiments at T2K~\cite{Abe:2013hdq} and $\bar{\nu}_e$ disappearance
experiments with reactor antineutrinos~\cite{An:2016bvr,RENO:2015ksa,Abe:2014bwa}
 have opened up the possibilities of studying physics of mass hierarchy(MH) and CP violation in the three flavor 
phenomenology of neutrino oscillations. The ongoing experiments at T2K and NO$\nu$A, and future experiments planned at DUNE, SBND and T2HK 
in appearance and disappearance channels with $\nu_\mu$ and $\bar\nu_\mu$ mode i.e. $\nu_\mu(\bar\nu_\mu) \to \nu_e(\bar\nu_e)$ and $\nu_\mu(\bar\nu_\mu) \to \nu_\mu(\bar\nu_\mu)$, respectively, 
will be able to determine various parameters of PMNS matrix with higher precision and answer decisively the question of neutrino mass hierarchy(MH) and CP violation in lepton 
sector.

In the neutrino oscillation experiments with $\nu_\mu(\bar\nu_\mu)$ beams in the appearance channels i.e. $\nu_\mu(\bar\nu_\mu) \to \nu_e(\bar\nu_e)$, the major source of background 
comes from the beam contamination due to presence of $\nu_e(\bar\nu_e)$ in $\nu_\mu(\bar\nu_\mu)$ beams. It is, therefore, extremely important to understand theoretically the difference between the 
interaction cross sections of various processes induced by $\nu_\mu(\bar\nu_\mu)$ and $\nu_e(\bar\nu_e)$ on nucleon and nuclear targets. In the present experiments at T2K, where the target 
material will mainly consists of $^{12}$C($^{16}$O) in near(far) detector, the nuclear effects will be quite important as the beam energy is low i.e. $E_\nu \sim$ 0.6 GeV. On the other hand at NO$\nu$A,
 where the beam energy is relatively higher i.e. $E_\nu \sim$ 2 GeV, the nuclear effects will be relatively smaller but still there will be significant differences in the reaction cross section 
 induced by $\nu_\mu(\bar\nu_\mu)$ and $\nu_e(\bar\nu_e)$ as discussed recently by Day and McFarland~\cite{Day:2012gb} in the case of quasielastic processes. Moreover, at NO$\nu$A energies,
 there would be 
 contribution from inelastic processes of one pion production and deep inelastic scattering(DIS) for which there have been no comparative studies of lepton production yields induced by 
 $\nu_\mu(\bar\nu_\mu)$ and $\nu_e(\bar\nu_e)$.
 
 Day and McFarland~\cite{Day:2012gb} have shown that the cross sections for charged current quasielastic processes induced by $\nu_\mu(\bar\nu_\mu)$ and $\nu_e(\bar\nu_e)$ would be different even in
 the presence of lepton universality of weak interaction due to various reasons which owe their origin to different masses of produced charged leptons $\mu^-(\mu^+)$ and $e^-(e^+)$, which are:
 \begin{itemize}
  \item Different kinematics for charged leptons $\mu^-(\mu^+)$ and $e^-(e^+)$ in presence of lepton mass will show in $\frac{d\sigma}{dQ^2}$, $\frac{d\sigma}{dE}$, $\frac{d\sigma}{d\theta}$ due to change 
  in kinematical limits of $Q^2$, E and $\theta$.
  \vspace{-2mm}
  \item The effects of uncertainties in vector and axial vector form factors which are quite important in the case of axial vector form factor in its use of axial dipole mass $M_A$ will affect the cross 
  section differently for $\mu^-(\mu^+)$ and $e^-(e^+)$ production due to change in kinematic variables in presence of lepton mass.
  \vspace{-2mm}
  \item There will be additional contribution due to pseudoscalar from factor which will be different for $\mu^-(\mu^+)$ and $e^-(e^+)$ production. This could be important at
  low $\nu_{_l}(\bar{\nu_{_l}})$ energies.
  \vspace{-2mm}
  \item In the presence of second class currents(SCC) in the phenomenology of weak currents, the contribution of SCC being lepton mass dependent will be different for $\mu^-(\mu^+)$ and $e^-(e^+)$ production.
  \vspace{-2mm}
  \item The effect of radiative corrections, being dependent on lepton mass will be different for $\mu^-(\mu^+)$ and $e^-(e^+)$ production.
 \end{itemize}
In the context of present and future experiments at T2K, NO$\nu$A, DUNE, SBND and T2HK which will be using nuclei like $^{12}$C, $^{16}$O, $^{40}$Ar, $^{56}$Fe and $^{208}$Pb 
as target materials, it is desirable that we have a quantitative estimate of nuclear medium effects on the total cross sections and other observables like the angular and energy distributions of charged 
leptons which are used in the experimental analysis with this goal in mind, we have extended the study of Day and McFarland~\cite{Day:2012gb} to various nuclear targets and calculated various
observables which 
can be measured in these experiments. In this contribution, we report our results due to the effect of considering various physics inputs mentioned in this section on the total cross section of quasielastic 
charged current reactions when nuclear medium effects are also included~\cite{Akbar:2015yda}.

We have performed our calculations in the local Fermi gas model(LFG) including the effect of Fermi motion and Pauli blocking. The effect of nucleon-nucleon correlations are also incorporated through
the interaction of particle--hole (1p--1h) excitation in the nuclear medium in a random phase approximation(RPA) following our earlier work~\cite{Singh:1992dc} and
work of Nieves et al.~\cite{Nieves:2004wx}. The present results are compared with 
the results available in some other versions of Fermi gas model~\cite{LlewellynSmith:1971zm,Gaisser:1986bv,Smith}. It would be interesting to extend this work to compute total lepton yield
due to inelastic and DIS processes induced by 
$\mu^-(\mu^+)$ and $e^-(e^+)$ in the presence of lepton mass effects in the region of E$_{\nu_{_l}(\bar\nu_{_l})}\sim$ few GeV.
 
\subsection{Formalism}
The basic reaction for the quasielastic process, where a (anti)neutrino interacts with a
(proton)neutron target is given as:
\begin{eqnarray}\label{rxn}
 \nu_l(k) + n(p) \to l^-(k^\prime) + p(p^\prime);\;\;\;\;\;\;\;
 \bar \nu_l(k) + p(p) \to l^+(k^\prime) + n(p^\prime)
\end{eqnarray}

Transition matrix element for reactions given Eq.~\ref{rxn} is:
\begin{equation}
 {\cal M} = \frac{G_F}{\sqrt2} \cos\theta_c~ l_\mu~ J^\mu
\end{equation}
 where $G_F$ is Fermi coupling constant and $\theta_c$ is Cabibbo angle.

 Leptonic weak current is given by:
\begin{equation}
 l_\mu= \bar{u}(k^\prime) \gamma_\mu (1\pm \gamma_5)u(k)
\end{equation}
 where (+)-ve sign is for (antineutrino)neutrino. Hadronic current is given by:
\begin{eqnarray}
J^\mu&=& \bar u(p^\prime)\left[F_1^V(Q^2)\gamma^\mu+F_2^V(Q^2)i\sigma^{\mu\nu}\frac{q_\nu}{2M} 
+ F_3^V(Q^2)\frac{q^\mu}{M}\right. \nonumber \\
&+& \left.F_A(Q^2)\gamma^\mu\gamma^5 
+ F_P(Q^2) \frac{q^\mu}{M}\gamma^5  + F_3^A(Q^2)\frac{(p+p^\prime)^\mu}{M}\gamma^5\right] u(p),
\end{eqnarray}
$Q^2(=-q^2)~\geq 0$ is the four momentum transfer square and $M$ is the nucleon mass. $F_{1,2}^V(Q^2)$
 are the isovector vector form factors and
 $F_A(Q^2)$, $F_P(Q^2)$ are the axial and pseudoscalar form factors, respectively. 
$F_3^V(Q^2)$ and $F_3^A(Q^2)$ are the form factors
related with the second class current.
Isovector vector form factors $F_{1,2}^V(Q^2)$ of the nucleons are given as
\begin{equation}\label{f1v_f2v}
F_{1,2}^V(Q^2)=F_{1,2}^p(Q^2)- F_{1,2}^n(Q^2) 
\end{equation}
where $F_{1}^{p(n)}(Q^2)$ and $F_{2}^{p(n)}(Q^2)$ are the Dirac and Pauli form factors of proton(neutron) 
which in turn are expressed in terms of the
experimentally determined Sach's electric $G_E^{p,n}(Q^2)$ and magnetic $G_M^{p,n}(Q^2)$ form factors.

Dipole form has been used for the axial form factor, $F_A(Q^2) = F_A(0)\left[1 + \frac{Q^2}{M_A^2}\right]^{-2},$ with $F_A(0)$ = --1.267 and world average(WA) value of axial dipole mass $M_A$= 1.026 GeV. 
Pseudoscalar form factor $F_P(Q^2)$, is obtained by using Goldberger--Treimann relation~\cite{LlewellynSmith:1971zm}:
\begin{equation}\label{fp}
F_P(Q^2)=\frac{2M^2F_A(Q^2)}{m_\pi^2+Q^2}.
\end{equation}

We have used the following expressions for $F_3^V(Q^2)$ and $F_3^A(Q^2)$ as given in Ref.~\cite{Day:2012gb}:
\begin{equation}\label{eq:SCC}
F_3^V(Q^2)=4.4~F_1^V(Q^2);\qquad F_3^A(Q^2) =0.15~F_A(Q^2). 
\end{equation}
 The double differential cross section on free nucleon is obtained as:
\begin{equation}
\sigma_{free}(E_l,\Omega_l)\equiv \frac{d^2 \sigma}{ d E_l \; d \Omega_l }=
\frac{{|\vec k^\prime|}}{64\pi^2 E_\nu E_n E_p }{\bar\Sigma}\Sigma{|{\cal M}|^2}\delta[q_0+E_n-E_p]
\end{equation}
 In the local density approximation(LDA), which takes into account Pauli blocking, Fermi motion with RPA correlations,
the cross section is given by(details are given in Ref.~\cite{Akbar:2015yda}):
 \begin{eqnarray}
\sigma(E_\nu)&=&-2{G_F}^2\cos^2{\theta_c}\int^{r_{max}}_{r_{min}} r^2 dr 
\int^{{k^\prime}_{max}}_{{k^\prime}_{min}}k^\prime dk^\prime 
\int_{Q_{min}^{2}}^{Q_{max}^{2}}dQ^2\frac{1}{E_{\nu}^2 E_l}L_{\mu\nu}{J^{\mu\nu}_{RPA}} \nonumber \\
&\times& Im{U_N}[E_{\nu} - E_l - Q_{r} - V_c(r), \vec{q}]
\end{eqnarray}
\begin{figure}
\begin{center}
  \includegraphics[height=4.5cm,width=10cm]{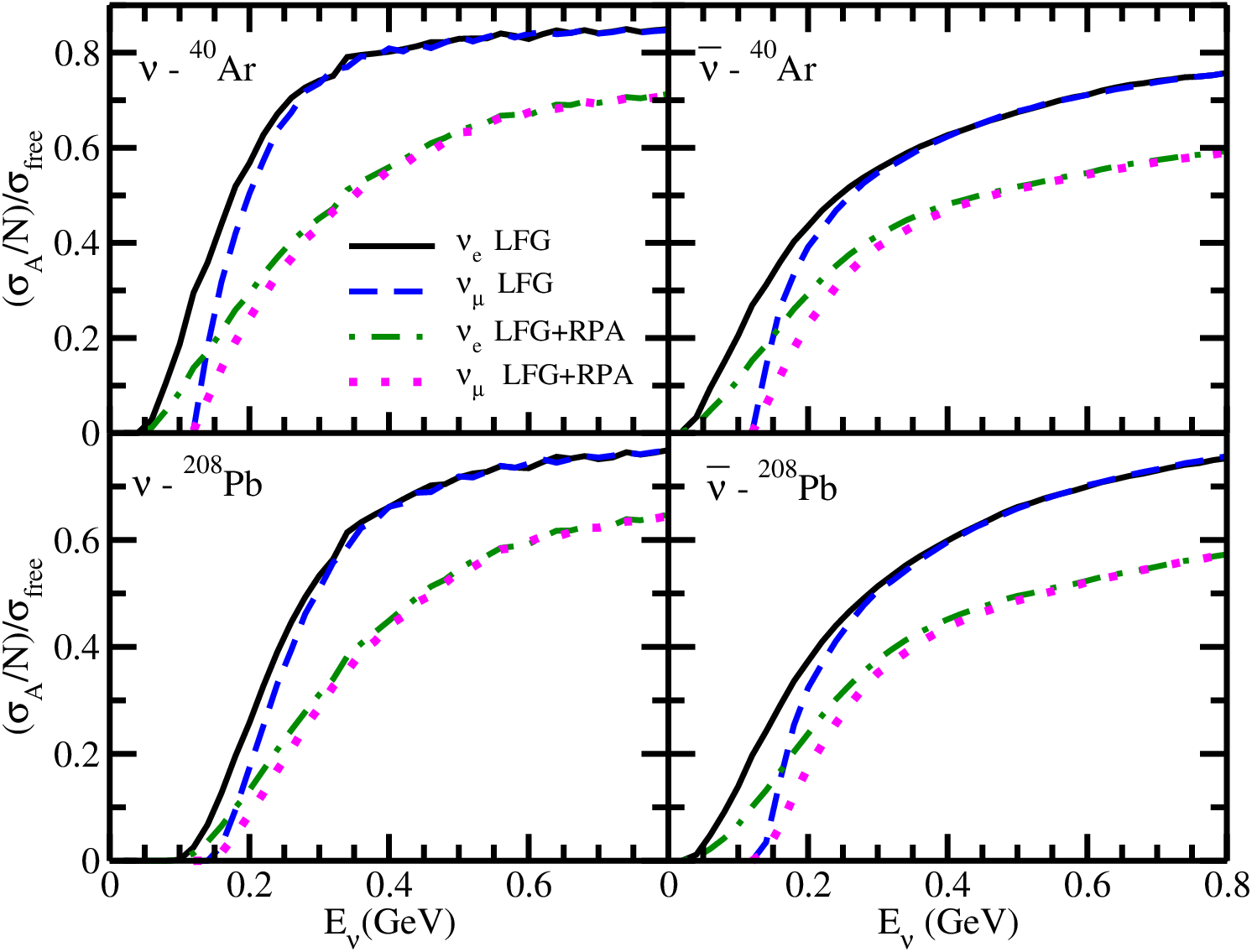}
 \caption{Ratio $\frac{\sigma_A/N}{\sigma_{free}}$ vs E$_\nu$, for neutrino(Left panel) and antineutrino(Right panel) induced processes in $^{40}$Ar and $^{208}$Pb. 
 The solid(dashed) line represent cross section obtained from
electron(muon) type neutrino and antineutrino beams. For neutrino induced process N = A -- Z,
is neutron number and for antineutrino induced process N = Z, is proton number. $\sigma_A$ is cross
section in nuclear target and has been evaluated using Local Fermi Gas Model(LFG) and LFG
with RPA effect(LFG+RPA) and $\sigma_{free}$ is the cross section for the free nucleon case.}\label{fig1}
\end{center}
\end{figure}
To study the lepton mass dependence on electron and muon type (anti)neutrino induced
scattering cross sections in free nucleon as well as nuclear target, we define:
\begin{equation}\label{delta_I}
 \Delta_{I}= \frac{\sigma_{\nu_e(\bar\nu_e)}-\sigma_{\nu_\mu(\bar\nu_\mu)}}{\sigma_{\nu_e(\bar\nu_e)}}
\end{equation}
 where I =(i) free (anti)neutrino-nucleon case,(ii)LFG,(iii)LFG+RPA effect.\\
We have studied the dependence of the cross section on axial dipole mass by using $\delta_{M_A}$, which is given as:
 \begin{equation}\label{sdelta}
 \delta_{M_A}=\frac{\sigma_{\nu_l}(M_A^{modified}) - \sigma_{\nu_l}(M_A= 
WA)}{\sigma_{\nu_l}(M_A= WA)}.
 \end{equation}
 To study the effect of pseudoscalar and second class vector form factors on the scattering
cross section, we define:
\begin{equation}\label{delta_F}
 \Delta_{F_i} = \left(\frac{{\sigma_{\nu_\mu}}(F_i \neq 0) - 
{\sigma_{\nu_e}}(F_i \neq 0)}{{\sigma_{\nu_e}}(F_i \neq 0)} \right) - \left( \frac{{\sigma_{\nu_\mu}}(F_i = 0) - {\sigma_{\nu_e}}(F_i= 0)}{{\sigma_{\nu_e}}(F_i = 0)}\right)
\end{equation}
 where $F_i$ stands for either $F_P(Q^2)$, $F_3^V(Q^2)$ or $F_3^A(Q^2).$
 
   To observe the effect of radiative corrections on the scattering cross section, we define
\begin{equation}\label{delta_RC}
 \Delta_{RC}=\left(\frac{{\sigma_{\nu_\mu}}^{RC} - {\sigma_{\nu_e}}^{RC}}{{\sigma_{\nu_e}}^{RC}} \right) - \left(\frac{{\sigma_{\nu_\mu}}^{NR} - 
{\sigma_{\nu_e}}^{NR}}{{\sigma_{\nu_e}}^{NR}} \right),
\end{equation}
where ${\sigma_{\nu_l}}^{RC}$ and ${\sigma_{\nu_l}}^{NC}$ stand for total cross section with and without radiative corrections.
% \vspace{-5mm}
\subsection{Results and discussion}
In Fig.~\ref{fig1}, we have shown the ratio of total nuclear cross section per interacting nucleon to total cross section for free nucleon target
for $\nu_l(\bar \nu_l)$ induced CCQE 
scattering in $^{40}$Ar and $^{208}$Pb separately 
for $\nu_e(\bar \nu_e)$ and $\nu_\mu(\bar \nu_\mu)$. To incorporate nuclear medium effects, we have used 
LFG with and without nucleon correlation effects using RPA . 
For $^{40}$Ar, we observe reduction in the total cross section when we include nuclear medium effects using LFG, for example at $E_{\nu}$ = 0.3 GeV the decrease is $\sim 26\%$($44\%$)
for $\nu_e(\bar \nu_e)$ and $26\%$($45\%$) for $\nu_\mu(\bar \nu_\mu)$. 
With the inclusion of RPA i.e LFG+RPA we observe further reduction. At $E_{\nu}$ = 0.3 GeV the reduction in the total cross section with LFG+RPA for $\nu_e(\bar \nu_e)$ and $\nu_\mu(\bar \nu_\mu)$ induced 
reactions is $\sim 55\%$($58\%$) and $57\%$($61\%$), respectively.

\begin{figure}
\begin{center}
  \includegraphics[height=4.5cm,width=10cm]{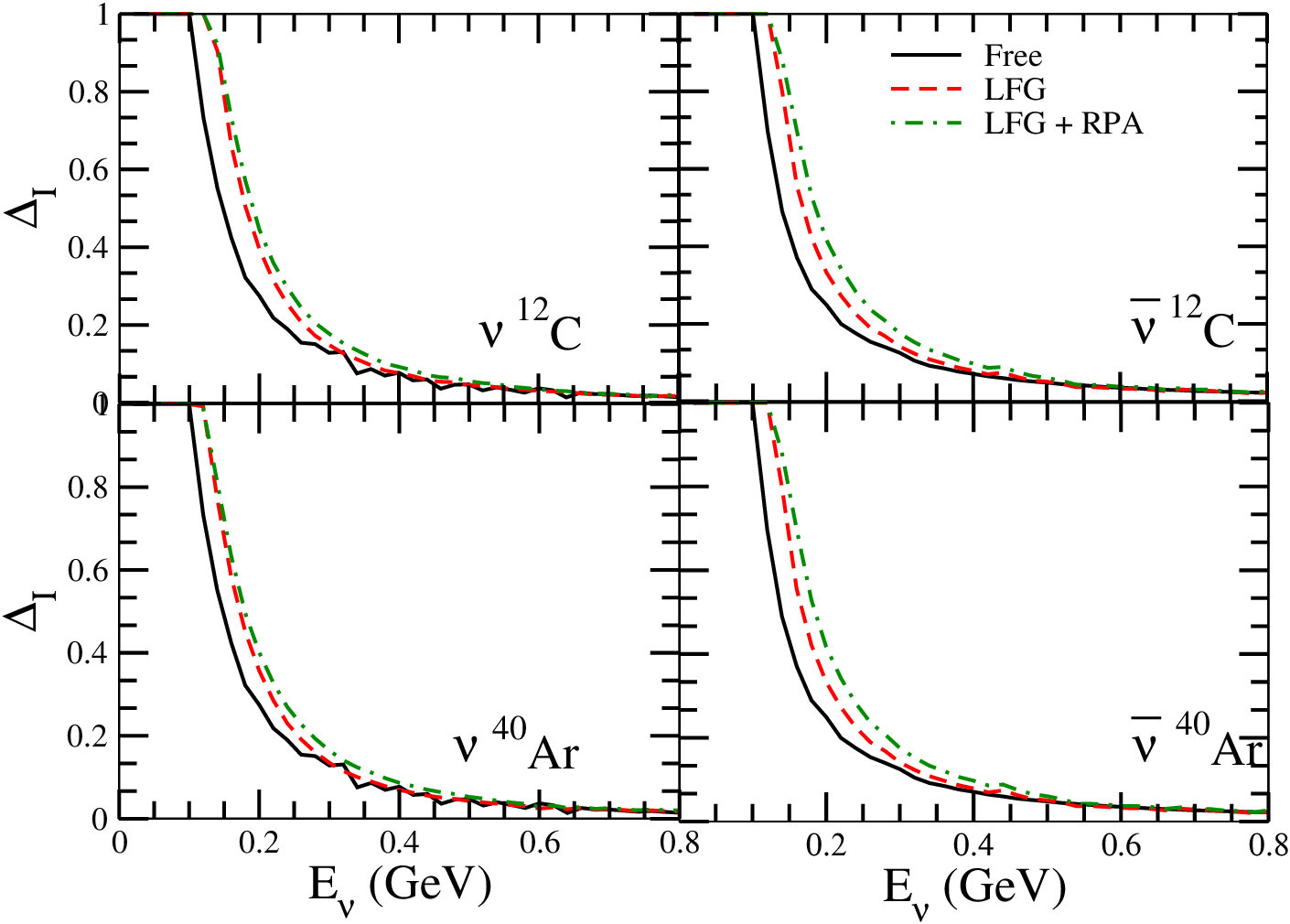}
 \caption{$\Delta_I$(Eq.~\ref{delta_I})
for neutrino(left panel) and antineutrino(right panel) induced processes in $^{12}$C and $^{40}$Ar targets. Here I stands for the results of the cross sections obtained (i) for
the free nucleon case(solid line) (ii) in the Local Fermi Gas Model(dashed line) and (iii) LFG
with RPA effect(dashed dotted line).}\label{fig2}
\end{center}
\end{figure}

In Fig.~\ref{fig2}, we have shown the effect of lepton mass by defining $\Delta_I$(Eq.~\ref{delta_I}) for (i) free nucleon case, (ii) LFG, and (iii) LFG + RPA for $^{12}$C and $^{40}$Ar nuclear targets. 
At low energies, the difference in the production cross section for $\nu_e(\bar \nu_e)$ and $\nu_\mu(\bar \nu_\mu)$ is large and vanishes with the increase in incoming (anti)neutrino energy. Also,
at low energies, as we go from free nucleon case to nuclear medium the reduction increases, for example at $E_\nu$ = 0.2 GeV the fractional change for $\nu_l(\bar \nu_l)$ induced reactions
is $\sim$ 27$\%$(25$\%$) in the case of free nucleon, while in $^{12}$C using LFG it is $\sim$ 40$\%$(33$\%$) and with the inclusion of RPA with LFG the reduction is $\sim$ 44$\%$(42$\%$).

% =============================================================================
% =============================================================================
\begin{figure}
\begin{center}
  \includegraphics[height=4.5cm,width=10cm]{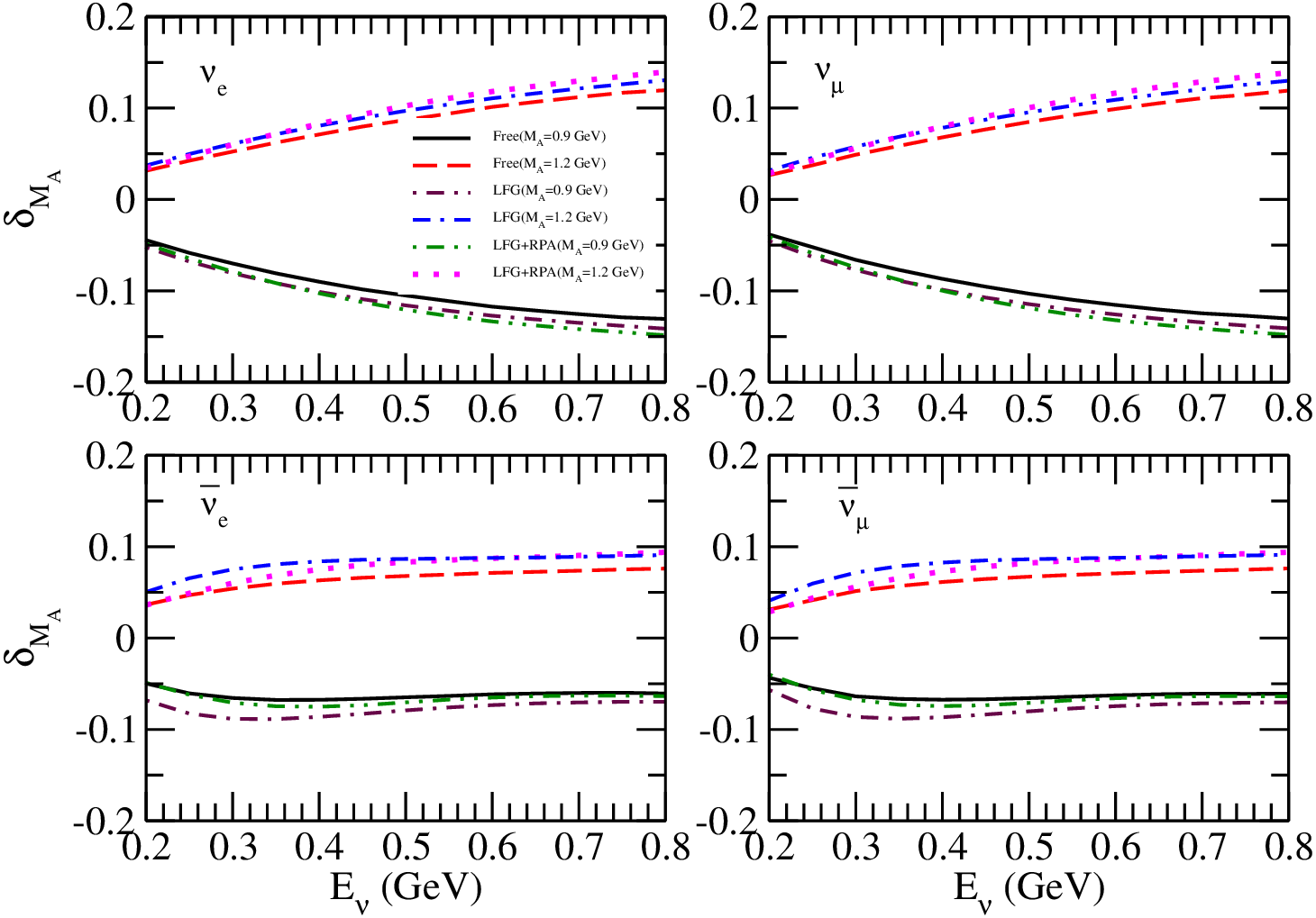}
 \caption{The dependence of cross section on $M_A$ obtained using Eq.~\ref{sdelta}. The results are shown
for $\nu_e(\bar\nu_e)$ and $\nu_{_\mu}(\bar\nu_{_\mu})$ induced processes on free nucleon as well as on $^{40}$Ar target using LFG
with and without RPA effect. Solid(dashed) line denotes results for the free nucleon case with
$M_A$ = 0.9 GeV (1.2 GeV ), results obtained using LFG are shown by dashed dotted(double dashed
dotted) and results for LFG with RPA effect are shown by dashed
double dotted(dotted).}\label{fig3}
\end{center}
\end{figure}
% \vspace{-5mm}
% =============================================================================
\begin{figure}
\begin{center}
  \includegraphics[height=4.3cm,width=10cm]{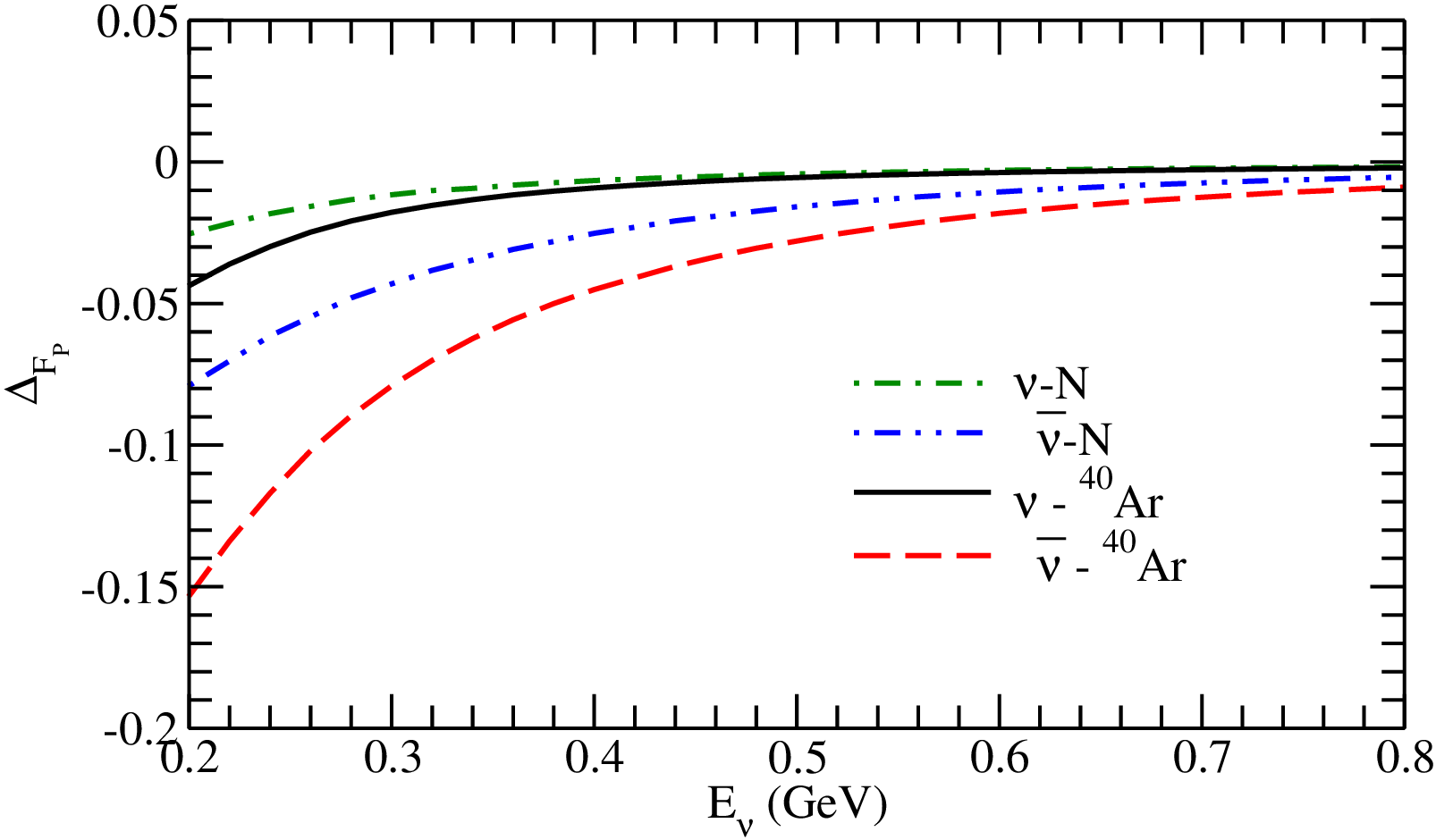}
 \caption{Results of the fractional change $\Delta_{F_P}$ defined in Eq.~\ref{delta_F} as a function of (anti)neutrino energy.
Results for $\nu$ induced cross section for free nucleon
case(dashed dotted line) and for $^{40}$Ar(solid line) 
obtained by using LFG with RPA effect. For $\bar \nu$, the results are shown by dashed double dotted line(free nucleon case) and dashed line($^{40}$Ar target).}\label{fig4}
\end{center}
\end{figure}
% =============================================================================
\begin{figure}
\begin{center}
  \includegraphics[height=4.3cm,width=10cm]{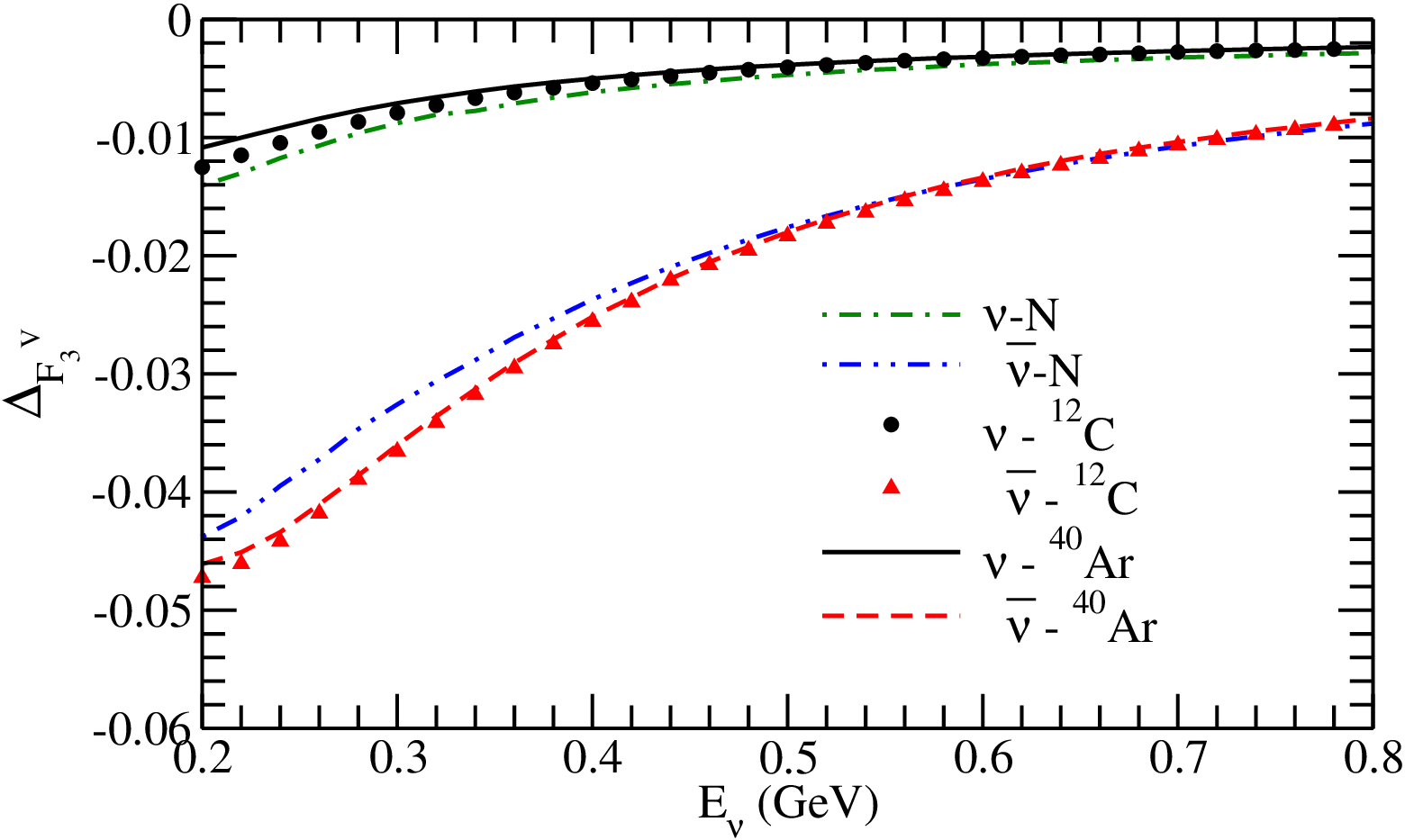}
 \caption{The difference of fractional changes $\Delta_{F^V_3}$
defined in Eq.~\ref{delta_F}, for the free nucleon
case($\nu$ results shown by dashed-dotted line and $\bar \nu$ results by dashed-double dotted line) as well as for $^{12}$C 
(circle for $\nu$ and triangle up for $\bar \nu$) and $^{40}$Ar (solid
line for $\nu$ and dashed line for $\bar \nu$) nuclear targets obtained by using LFG.}\label{fig5}
\end{center}
\end{figure}
% ===================================================================================================
% \vspace{-5mm}
\begin{figure}
\begin{center}
  \includegraphics[height=4.3cm,width=10cm]{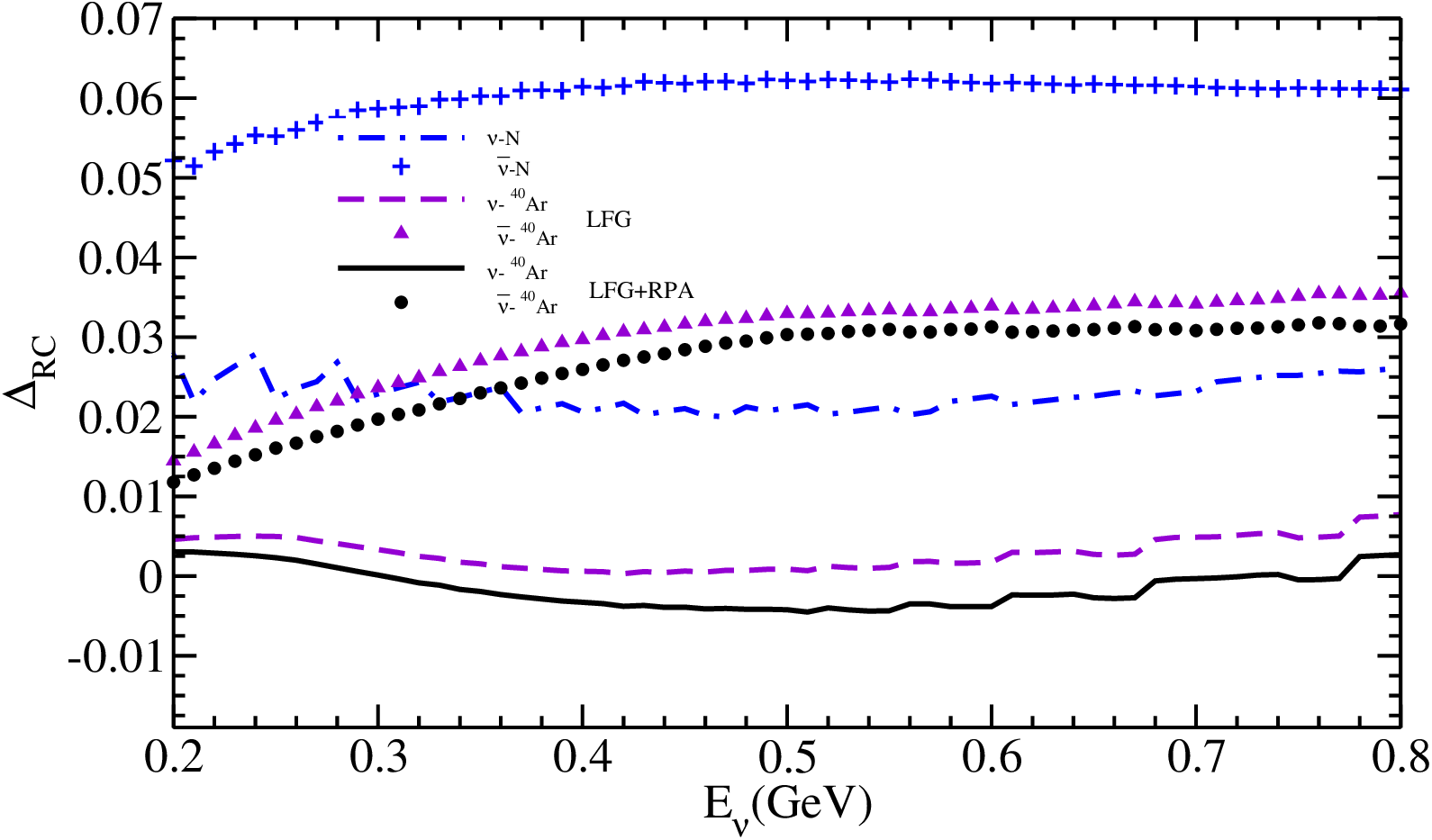}
 \caption{The effect of radiative corrections on fractional difference $\Delta_{RC}$ defined in Eq.~\ref{delta_RC} for
(anti)neutrino induced processes on free nucleon as well as on $^{40}$Ar target using LFG with
and without RPA effect. The results are shown for the $\nu(\bar \nu)$ induced processes on free nucleon by dashed dotted line(plus), for LFG by dashed line (triangle up)
and for LFG with RPA effect by solid line(circle).}\label{fig6}
\end{center}
\end{figure}
We have studied the dependence of total cross section on the axial dipole mass $M_A$ for free nucleon case as well as in $^{40}$Ar nuclear target. 
The numerical results are presented in Fig.~\ref{fig3}, where we have plotted $\delta_{M_A}$(Eq.~\ref{sdelta}).
 $M_A$ variation is shown by taking 1.026 GeV as reference value and modified values to be 0.9 GeV and 1.2 GeV.
% We have shown the variation by taking $M_A$ as 1.026 GeV(reference value) and 0.9 GeV and 1.2 GeV.

The effect of pseudoscalar form factor on the production cross section of electron and muon for both free nucleon and nucleons bound in $^{40}$Ar 
 nuclear target is obtained by using LFG with RPA. 
The results are presented in Fig.~\ref{fig4} by using $\Delta_{F_P}$ defined in Eq.~\ref{delta_F}. We observe large fractional change in case of 
 antineutrino induced cross section as compared to neutrino 
induced cross section when we go from free nucleon case to nuclear target. Moreover, with the increase in $E_\nu$,
the difference vanishes for both $\nu_l$ and $\bar \nu_l$.

In Fig.~\ref{fig5},  we present the contribution of $F_3^V(Q^2)$ to the total scattering cross section for free nucleon case and in $^{12}$C and $^{40}$Ar nuclear targets. 
Since $F_3^V(Q^2)$ is proportional to the lepton mass, therefore, the dependence will be different in $\nu_e(\bar \nu_e)$ case from $\nu_\mu(\bar \nu_\mu)$ case. At low (anti)neutrino energies
we observe small 
effect of $F_3^V(Q^2)$ as we move from free nucleon case to nuclear targets. From the
figure it may be noticed that the fractional change is the same for both $^{12}$C and $^{40}$Ar nuclei.

 To see the effect of radiative corrections, we use Eq.~\ref{delta_RC} to define $\Delta_{RC}$ and present our numerical results in Fig.~\ref{fig6} 
for free nucleon as well as on nuclear targets using LFG 
with and without RPA. We observe small dependence of total cross section on the radiative corrections for antineutrino induced reactions in the case of free nucleon and nuclear targets.  

Thus we find that nuclear medium effects are different in antineutrinos than in neutrinos. Due to threshold effect the suppression in the cross section shows 
a different behavior for $\nu_e(\bar \nu_e)$ and $\nu_\mu(\bar \nu_\mu)$ induced processes for $E_\nu~<~0.4GeV$. The effect of radiative corrections is more pronounced in 
 electron events than in muon events. 
% =============================================================================

% =============================================================================

\end{document}